\documentclass[aps,prb,preprint,superscriptaddress,amsmath,amssymb,showpacs,superscriptaddress]{revtex4-1}

\usepackage{graphicx}
\usepackage{dcolumn}
\usepackage{bm}
\usepackage{color}
\usepackage{tikz}
\usetikzlibrary{plotmarks}
\usepackage{epstopdf}
\usepackage{amssymb}

\begin{document}



\title{Telegraph frequency noise in electromechanical resonators}


\author{F. Sun}
\affiliation{Department of Physics, the Hong Kong University of Science and Technology, Hong Kong, China }

\author{J. Zou}
\affiliation{Department of Physics, the Hong Kong University of Science and Technology, Hong Kong, China }
\affiliation{Department of Physics, University of Florida, Gainesville, Florida 32611, USA }

\author{Z. A. Maizelis}
\affiliation{I.M. Lifshyts Department of Theoretical Physics, Kharkiv National University, Kharkov 61022, Ukraine}
\affiliation{A. Ya. Usikov Institute for Radiophysics and Electronics, National Academy of Sciences of Ukraine, Kharkov 61085, Ukraine}

\author{H. B. Chan}
\email[]{hochan@ust.hk}
\affiliation{Department of Physics, the Hong Kong University of Science and Technology, Hong Kong, China }


\begin{abstract}
We demonstrate experimentally the possibility of revealing fluctuations in the eigenfrequency of a resonator when the frequency noise is of the telegraph type. Using a resonantly driven micromechanical resonator, we show that the time-averaged vibration amplitude spectrum exhibits two peaks. They merge with an increasing rate of frequency switching and the spectrum displays an analog of motional narrowing. We also show that the moments of the complex amplitude depend strongly on the frequency noise characteristics. This dependence remains valid even when strong thermal or detector noise is present.
\end{abstract}


\maketitle

\section{INTRODUCTION}

The decay and loss of coherence of oscillations is of paramount importance in resonators \cite{Gitterman2005noisy}, from Josephson junctions \cite{Clarke2008,LiJian2013} to nano- and optomechanical systems \cite{Cleland2002, Kippenberg2008,Faust2013,Okamoto2013}. In the vast majority of cases where the resonator is coupled to a thermal bath, the decay of vibrations is associated with thermal fluctuations in both the amplitude and the phase of the vibrations. Phase noise is particularly important because it directly determines the sensitivity of these devices to act as resonant detectors \cite{Giessibl2003} and their stability to serve as precision clocks \cite{Sullivan1990}. Apart from originating thermally, phase fluctuations can also arise due to fluctuations in the eigenfrequency of the resonator \cite{Rubiola2007}. In mechanical systems, such frequency noise can be generated by a number of sources, including the coupling of the vibration mode to two-level systems \cite{Rugar2004, LaHaye2009}, random adsorption and desorption of molecules on the surface of the resonator \cite{Jensen2008,Wang2010,Hanay2012}, as well as the diffusion of adsorbed molecules along the resonator \cite{Yang2011}. 

There has been much interest in identifying frequency noise because the study of its statistics allows one to probe the underlying physics of mesoscopic systems \cite{Ekinci2004,Hanay2012,Clerk2010a}. Frequency noise can be directly measured if the eigenfrequency varies slowly with time and thus can be accurately measured by, for instance, fitting to a resonance peak. Rapid changes of the eigenfrequency (faster than the inverse damping constant) may be detectable by observing the ringdown of oscillations or driving the resonator into self oscillations with a phase locked loop. However, due to the presence of detector noise and/or thermal noise, in practice it is not straightforward to isolate frequency fluctuations from other sources of phase noise. 

To characterize phase noise, a common method involves driving the resonator with a sinusoidal excitation at a frequency close to the eigenfrequency and measuring the two oscillation quadratures $X$ and $Y$ that are in phase and out of phase with the drive \cite{Sullivan1990,Gao2007,Fong2012}. For sufficiently large oscillation amplitudes, the anisotropy of the distribution in the $X$-$Y$ phase space due to frequency noise can exceed the broadening due to thermal noise and detector noise. Using this method, the intrinsic frequency fluctuations of a silicon nitride resonator were recently measured \cite{Fong2012}. The feasibility of this approach relies on the resonator remaining in the linear regime even in the presence of a strong periodic drive. For resonators with a low threshold of the onset of nonlinearity, including graphene and carbon nanotube mechanical resonators \cite{Sazonova2004, Jensen2008, Bunch2007, Eichler2011}, such a scheme may not be applicable. Recently, it was proposed that frequency noise can be identified in the spectra of the higher moments of the complex amplitude \cite{Dykman2010a,Maizelis2011} of vibrations. Furthermore, the higher moments also provide useful information about noise characteristics. 

In the present paper, we perform a comprehensive study of telegraph frequency noise. The experiment is done with a micromechanical torsional resonator. When the resonator is subjected to random telegraph frequency noise (RTFN), its eigenfrequency jumps randomly between two values, $\omega_{1,2}= \omega_0 \pm \Delta/2$. The separations between the two frequencies $\Delta$ and the jumping rate $W$ are chosen to be much smaller than the eigenfrequency. We record the complex amplitude of oscillations $u$ in response to a sinusoidal driving force. When $\Delta$ is larger than the damping constant $\Gamma$, the amplitude spectrum $|\langle u \rangle|$ consists of two distinct peaks for small $W$ ($\langle \ldots \rangle$ stands for the time-averaged value). As $W$ increases, the two peaks merge into a single peak, displaying spectral broadening, followed by narrowing in a manner analogous to motional narrowing \cite{Anderson1953,Anderson1954}. For $\Delta\sim\rm{max}(\Gamma, W)$, the broadening is small, making it difficult to identify the existence of frequency noise solely from the conventional spectrum of the response to the periodic drive. We demonstrate that the deviation of the higher moments of the complex amplitude $\langle u^n \rangle$ from the powers of the averaged complex amplitude $\langle u \rangle^n$ unambiguously indicates the presence of frequency noise. The shapes of the spectra of $|\langle u^2 \rangle/\langle u \rangle^2|$ and $|\langle u^3 \rangle/\langle u \rangle^3|$ depend strongly on $W$ and $\Delta$ of the RTFN, in good agreement with theory. Remarkably, this method of studying the characteristics of the frequency noise remains valid even in the presence of strong thermal or detector noise, when conventional schemes fail to reveal the existence of frequency noise. It does not require driving the resonator to large vibrational amplitudes, and thus avoids the problems related to resonator nonlinearity.

\begin{figure}[h]
\includegraphics[width=8.6cm]{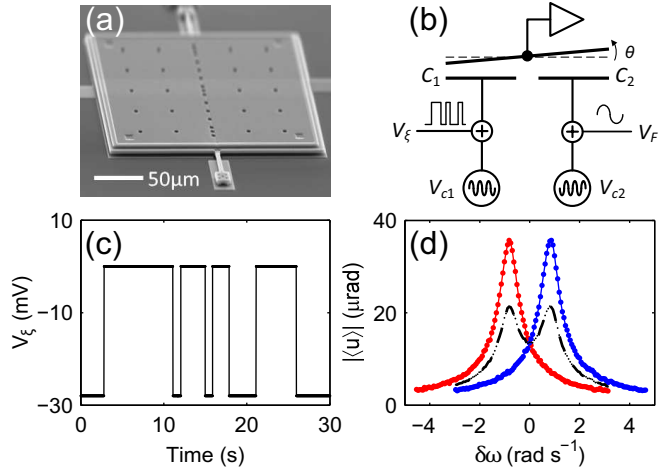}
\caption{\label{fig:1} (Color online) (a) Scanning electron micrograph of the micromechanical torsional oscillator. (b) Cross-sectional schematic of the device and the measurement circuitry. (c) Typical telegraph noise voltage that is applied to the left electrode. (d) The amplitude spectra of the resonator $|\langle u\rangle|$ with eigenfrequency shifts $\xi = \pm \Delta/2$ (right and left peak, respectively). $\Delta = 1.665$ rad s$^{-1}$ is constant in time. The dashed-dotted curve is an average of the two solid curves. 
}
\end{figure}

\section{MICROMECHANICAL TORSIONAL RESONATOR AND DETECTION CIRCUITRY}

The torsional resonator in our experiment consists of a highly doped polycrystalline silicon plate with dimensions $200~\mu$m~$\times~200~\mu$m$~\times 3.5 ~\mu$m, suspended by two torsional rods ($36~\mu$m~$\times~4~\mu$m$~\times~2 ~\mu$m) at the middle of the opposite edges [Fig.$\:$\ref{fig:1}(a)]. Two electrodes are located underneath the plate, one on each side of the torsional rods. The gap between the top plate and the electrodes is created by etching away a $2 ~\mu$m sacrificial silicon oxide layer. 

The rotation angle $\theta$ of the plate is measured by a capacitance bridge \cite{Chan2001a} that detects the difference of the capacitances $C_1$ and $C_2$ between the top plate and the two bottom electrodes [Fig.$\:$\ref{fig:1}(b)]. Two $180^{\circ}$ out-of-phase ac voltages ($V_{c1}$ and $V_{c2}$) at frequency $\omega_{\rm{carrier}}$ (typically $2.5\times10^7$ rad s$^{-1} \gg \omega_0$) are applied to the two bottom electrodes, respectively. As the plate rotates, $C_1$ increases and $C_2$ decreases. For small rotations, the amplitude of the ac voltage at $\omega_{\rm{carrier}}$ on the top plate is proportional to $C_1-C_2$, which is in turn proportional to $\theta$. A lock-in amplifier referenced to $\omega_{\rm{carrier}}$ measures the ac voltage on the top plate, yielding an output that is proportional to $\theta(t)$. All measurements are performed at $77$ K and $< 10^{-5}$ Torr. 

The equation of motion of the top plate is given by
\begin{equation}
\ddot{\theta}+2\Gamma \dot{\theta}+[\omega_0+\xi(t)]^2 \theta = F \cos (\omega_F t) + f_{\rm{th}}(t). 
\label{eq:Motion}
\end{equation}
The term $F \cos (\omega_F t)$ represents a periodic electrostatic torque generated by a sinusoidal voltage $V_F$ (amplitude $100~\mu$V) on top of a dc voltage ($-0.1$ V) applied to the electrode on the right. $\omega_F$ is chosen to be close to the eigenfrequency $\omega_0 = 134~024.383$ rad s$^{-1}$ in the absence of frequency noise $\xi(t)$. $F$ is kept small so that the oscillations remain linear. $\Gamma = 0.35$ rad s$^{-1}$ is the damping constant and $f_{\rm{th}}$ represents the thermal noise. To measure the oscillation amplitude, the output of the first lock-in amplfier [proportional to $\theta(t)$, as described earlier] is fed into a second lock-in amplifier that is referenced to $\omega_F$. The output of the second lock-in amplifier \emph{X(t)} and \emph{Y(t)} gives the oscillation amplitude in phase and out of phase with the periodic driving torque, respectively.
 
Changes in the resonant frequency $\xi(t)$ ($\ll \omega_0$) are induced by a voltage $V_{\xi}$ applied to the left electrode, by the spring softening effect associated with the electrostatic force gradient. Figure \ref{fig:1}(d) shows the oscillation amplitude as a function of $\delta \omega = \omega_F - \omega_0$. The blue and red curves are measured when $\xi$ remains constant with time, at values of $\pm\Delta/2$ where $\Delta =1.665$ rad s$^{-1}$, corresponding to dc voltages of $V_1 = 0$ V and $V_2 =-28$ mV, respectively, applied to the left electrode. 

Next, we replace the constant dc voltages by a voltage signal that consists of telegraph noise switching back and forth between the two voltages \emph{$V_1$} and \emph{$V_2$}, so that the eigenfrequency of the resonator jumps randomly between the two values $\omega_0 \pm \Delta/2$. The telegraph voltage signal is generated by a \emph{J-K} flip flop, triggered by Gaussian noise that is created by amplifying the Johnson noise of a $50~\Omega$ resistor. Switching of the flip flop occurs when the Gaussian noise exceeds a threshold. The rate of switching \emph{W} is therefore controlled by the Gaussian noise intensity. For all measurement presented here, \emph{W} is chosen to be at least $296$ times smaller than the eigenfrequency. We checked that the time intervals between switching of the flip flop are random and obey Poisson statistics. To eliminate ringing of the top plate, an \emph{RC} filter is used to increase the rise time of the voltage steps to $\tau = 0.011$ ms ($1/\omega_0 < \tau\ll 1/W$).

\section{COMPLEX VIBRATION AMPLITUDE AND THE HIGHER MOMENTS IN THE PRESENCE OF TELEGRAPH FREQUENCY NOISE}

In the analysis, it is convenient to transform the resonator motion into the rotating frame and introduce the complex vibrational amplitude $u = X + iY$ that is given by:
\begin{eqnarray}
\theta(t)=u\exp(i\omega_F t) + u^*\exp(-i\omega_F t),\nonumber\\
\frac{d\theta(t)}{dt}=i \omega_F [u\exp(i\omega_F t) - u^*\exp(-i\omega_F t)].
\end{eqnarray}
In the experiment, $X(t)$ and $Y(t)$ are the oscillation amplitudes that are in phase and out of phase with the periodic driving torque, respectively, measured by the second lock-in amplifier. The effect of different types of frequency noise on the time-averaged $|\langle u\rangle |$  and the higher moments $|\langle u^n\rangle |$  ($n = 1,2,\ldots$) was described in detail in 
Ref. [25].
Here we apply this theory to the telegraph frequency noise used in our experiment. $|\langle u^n\rangle |$ can be obtained from a vector $\mathbf {U}(n)$ with two components $U_i(n)$ ($i=1,2$),
\begin{eqnarray}
\mathbf M(n) \mathbf{U}(n)=-\frac{inF}{4\omega_F}\mathbf{U}(n-1),
\label{eq:moments}
\end{eqnarray}
where
\begin{eqnarray}
\mathbf M(n)=n(\Gamma+i\delta\omega)\mathbf{I}-\frac{in\mathbf{\sigma}_z\Delta}{2}+ \mathbf{W},\nonumber\\
\mathbf{W}=\frac{1}{2}\left(
       \begin{array}{cc}
         W & -W \\
         -W & W \\
       \end{array}
     \right),\quad\mathbf U(0)=\left(
       \begin{array}{cc}
         1\\
         1\\
       \end{array}
     \right).\nonumber
\end{eqnarray}
$\mathbf{I}, \mathbf{\sigma}_z$ are the identity matrix and Pauli matrix,  respectively. The solution of Eq.~(\ref{eq:moments}) is
\begin{equation}
\mathbf U(n)=n!\left( \frac{-iF}{4\omega_F}\right) ^n{\prod_{k=n}^1}\mathbf M^{-1} (k)\mathbf U(0).
\label{eq:sol}
\end{equation}
The averaged moments are given by $\langle u ^n\rangle=(1~~1)\cdot \mathbf U(n)/2$.

In the absence of frequency noise, $|\langle u\rangle |$ is the square root of a Lorentzian function [Fig.$\:$\ref{fig:1}(d)]. When the resonator is subjected to RTFN, the most intuitive case involves $W/\Delta \ll 1$, where either the interval between the two frequencies is large or the switching rate is small. In a simple picture, the resonator settles into one of the two oscillation states with well-defined amplitude and phase between consecutive switching events. One expects that the amplitude spectrum is the weighted summation of two independent amplitude spectra, with the weights being proportional to the occupation probability of the two states (both are $1/2$ for the telegraph noise we used). In Fig.$\:$\ref{fig:1}(d), the dashed-dotted curve is the average of the red and blue curves. Since $\Delta \sim 4.8 \Gamma$, two peaks can be clearly resolved. As $W/\Delta$ increases, the spectrum of $|\langle u\rangle |$ evolves according to Eq.~(\ref{eq:sol}) in a nontrivial manner.

\begin{figure}[h]
\includegraphics[width=8.6cm]{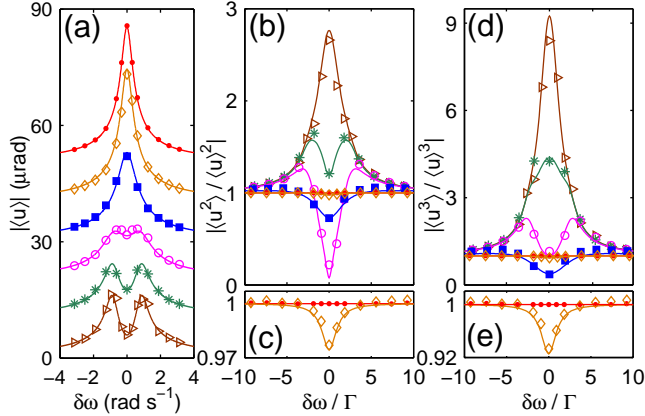}
\caption{\label{fig:2} (Color online) (a) Measured spectra of the time-averaged complex amplitude $|\langle u\rangle |$ of the resonator under RTFN with different $W/\Delta$. $\Delta$ is kept constant at $1.665$ rad s$^{-1}$. For clarity, successive curves are shifted by $10~\mu$rad. (The symbols
$\triangleright$, $\ast$, $\circ$, $\blacksquare$, $\diamond$, and $\bullet$ correspond to $W/\Delta$ = 0.05, 0.15, 0.45, 2.36, 43.26 rad$^{-1}$ and no frequency noise applied, respectively.) (b) The corresponding ratios $|\langle u^2\rangle /\langle u\rangle ^2| $ and (d) $|\langle u^3\rangle /\langle u\rangle ^3| $. (c) The expanded views of the ratios $|\langle u^2\rangle /\langle u\rangle ^2| $ and (e) $|\langle u^3\rangle /\langle u\rangle ^3| $ around 1. In all panels, the lines are the calculated spectra.
}

\end{figure}

\section{MEASURED COMPLEX VIBRATION AMPLITUDE: MERGING OF SPECTRAL PEAKS WITH INCREASING SWITCHING RATE}

Figure \ref{fig:2}(a) shows the measured $|\langle u\rangle |$  versus driving frequency $\omega_F$ when the resonator is subjected to RTFN with a fixed $\Delta$ but different switching rates $W$. The measurement time for each frequency is chosen to be sufficiently long (involving more than 300 switches) to accurately determine time-averaged values of $u$ . At $W/\Delta = 0.05$ rad$^{-1}$, two peaks are clearly resolved as expected [see the open triangles in Fig.$\:$\ref{fig:2}(a)]. However, as $W$ increases, the simple picture described above is no longer valid. For $W/\Delta = 2.36$ rad$^{-1}$, the two peaks merge into a single peak with a width smaller than $\Delta$ (solid squares). As predicted in 
Refs. [24,25],
the partial spectra are strongly coupled and interference occurs. An intuitive understanding is that in order to resolve two peaks separated by $\Delta$ by sweeping the driving frequency, one needs to measure for a duration longer than $1/\Delta$. When $W \gg \Delta, \Gamma$, the resident time in one state is not long enough for the resonator to discriminate between the two states and the spectrum becomes a single peak centered at $\omega_0$ whose width is broadened from the linewidth $\Gamma$ in the absence of frequency noise. As a result, the frequency noise leads to diffusion of the oscillator phase. As $W/\Delta$ is further increased to $43.26$ rad$^{-1}$, the single peak continues to narrow [open diamonds in Fig.$\:$\ref{fig:2}(a)]. In this limit, the resonator can no longer respond to the rapid changes in the resonant frequency. Vibrations occur as if the resonant frequency is held at a constant, time-averaged value. The scenario is analogous to motional narrowing in nuclear magnetic resonance where atoms undergo diffusion in liquids \cite{Anderson1953,Anderson1954}. When the spectral modulation amplitude is much smaller than the inverse of the modulation correlation time, the atoms experience an averaged magnetic field and  the resonance linewidth is smaller compared to the case when the atoms are stationary. In Fig.$\:$\ref{fig:2}(a), the lines are the calculated amplitude spectrum based on Eqs. (\ref{eq:moments}) and (\ref{eq:sol}) with no fitting parameters. There is good agreement between measurement and theory for all values of $W/\Delta$.

\section{MEASURED HIGHER MOMENTS OF THE COMPLEX VIBRATION AMPLITUDE}

Next, we study the effect of the RTFN on the averaged second and third moments. Figures \ref{fig:2}(b) and \ref{fig:2}(d) plot the ratios  $|\langle u^2\rangle /\langle u\rangle ^2| $ and $|\langle u^3\rangle /\langle u\rangle ^3| $, respectively, corresponding to the same parameter values in Fig.$\:$\ref{fig:2}(a). Figures \ref{fig:2}(c) and \ref{fig:2}(e) zoom into the regions where the ratios are close to 1. As shown by the solid circles, in the absence of external frequency noise, both ratios are equal to 1 within experimental uncertainty. In the presence of RTFN, they deviate from 1. The shape of the spectra is determined by $\Delta$ and $W$. At $W/\Delta = 0.05$ rad$^{-1}$, the spectrum of $|\langle u^2\rangle /\langle u\rangle ^2| $ displays a single peak centered at $\omega_0$ while the entire curve is above 1 (open triangles). As $W$ increases, the spectrum dips below $1$ with two maxima on its sides. When $W$ increases beyond $\Delta$ and $\Gamma$, the ratio approaches $1$. The spectra of $|\langle u^3\rangle /\langle u\rangle ^3|$ evolve in a qualitatively similar fashion. There is good agreement with theory (solid lines) in Figs.$\:$\ref{fig:2}(b)-2(e). 

\begin{figure}[h]
\includegraphics[width=8.6cm]{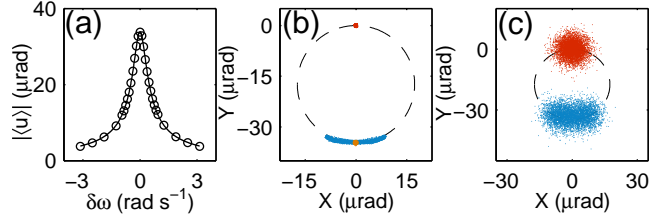}
\caption{\label{fig:3} (Color online) (a) Measured amplitude spectrum under weak frequency noise ($\Delta = 0.5 \Gamma$, $W/\Delta = 1.76$ rad$^{-1}$). The solid line is a fit of the square root of a Lorentzian, yielding a width that is $4.6\%$ larger than $\Gamma$. (b) The distribution of the measured complex amplitudes in the $X$-$Y$ phase diagram. With the external frequency noise removed, thermal motion of the resonator leads to isotropic distributions centered at the origin (no periodic drive, red) and at the bottom (periodic drive at $\omega_F = \omega_0$, orange). The dashed line represents the calculated $X$ and $Y$ when the driving frequency is swept through $\omega_0$. When the resonator is subjected to frequency noise as in (a), the distribution (blue) elongates along the dashed line. (c) The bandwidth of the lock-in is increased by about $1000$ so that detector noise makes it more difficult to identify frequency fluctuations.
}
\end{figure}

\section{EFFECTS OF THERMAL NOISE, DETECTOR NOISE AND MEASUREMENT TIME}

For several parameter values in Fig.$\:$\ref{fig:2}, the spectrum of  $|\langle u\rangle|$ exhibits a single peak in the presence of RTFN while $|\langle u^2\rangle /\langle u\rangle ^2|$ and $|\langle u^3\rangle /\langle u\rangle ^3| $ deviate significantly from $1$. In such cases, the analysis of the high moments of $u$ allows unambiguous identification of the existence of frequency noise even when thermal noise or detector noise is strong. Figure \ref{fig:3}(a) plots the averaged amplitude spectrum of the resonator for relatively weak frequency noise ($\Delta = 0.5 \Gamma$ and $W/\Delta = 1.76$ rad$^{-1}$). The spectrum is slightly broadened but is still well fitted by the square root of a Lorentzian, yielding a width that is $4.6\%$ larger than the intrinsic damping constant $\Gamma$. By examining the width of the amplitude spectrum alone, it is not possible to separate the effects of frequency noise from damping. To reveal the existence of frequency noise, the conventional method involves examining the complex amplitude distribution in the \emph{X}-\emph{Y} phase space. Figure \ref{fig:3}(b) shows three different distributions when $\omega_F$ is set to be equal to $\omega_0$. For the red points, the driving amplitude $F$ is turned to zero and external frequency noise is absent. Readings of the lock-in amplifier are recorded every $33$ ms with a time constant of $300$ ms. The finite width of the distribution is associated with the thermal mechanical fluctuations ($\sim$ 0.2 $\mu$rad) of our resonator at $77$~K. When $F$ is turned back on, the distribution is displaced along the vertical axis so that it is centered at a negative \emph{Y} value, because the average oscillation phase lags behind the drive by $\pi/2$ at resonance. Both distributions are circular in shape due to the phase-random nature of thermal noise. When the RTFN is introduced, the distribution becomes elongated along the \emph{X} direction. Such anisotropic broadening is distinct from that associated with thermal noise. In general, it serves as a signature for the presence of frequency noise. 

The aforementioned method of isolating the contributions of frequency noise becomes inapplicable when the thermal noise of the resonator or detector noise in the electronic circuit is large. In Fig.$\:$\ref{fig:3}(c), we repeat the measurement with the detection bandwidth increased by about a factor of $1000$ while the external frequency noise remains unchanged. The phase-random detector noise leads to additional isotropic broadening in the \emph{X}-\emph{Y} phase space, beyond that induced by thermal noise. For the distribution in Fig.$\:$\ref{fig:3}(c) in the presence of frequency noise and periodic driving, the elongation in the distribution due to frequency noise is largely masked by the detector noise. Figure \ref{fig:3}(c) shows that it becomes significantly more difficult to identify frequency noise by inspecting the distribution in the \emph{X}-\emph{Y} phase space when either the thermal noise or the detector noise is large.

\begin{figure}[h]
\includegraphics[width=8.6cm]{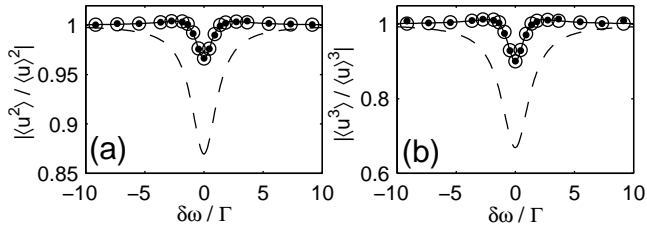}
\caption{\label{fig:4} Spectra of the ratios (a) $|\langle u^2\rangle /\langle u\rangle ^2|$ and (b) $|\langle u^3\rangle /\langle u\rangle ^3|$ for the same experimental conditions and under the same frequency noise as in Fig.$\:$\ref{fig:3}. The open and solid circles correspond to lock-in time constants of $300$ ms and $300~\mu$s, respectively. They correspond to (b) and (c) of Fig.$\:$\ref{fig:3}, respectively. The solid line is the theoretical prediction using Eq. (\ref{eq:sol}). The dashed line shows the calculated values when the telegraph frequency noise is replaced by white Gaussian frequency noise with the intensity $D=\Delta/2\sqrt{2}$.
}
\end{figure}

Figure \ref{fig:4}(a) plots, for the same experimental conditions and under the same frequency noise as in Fig.$\:$\ref{fig:3}, the spectra of the ratio $|\langle u^2\rangle /\langle u\rangle ^2|$. The solid circles and open circles in Fig.$\:$\ref{fig:4}(a) correspond to the measured values for large [Fig.$\:$\ref{fig:3}(c)] and small [Fig.$\:$\ref{fig:3}(b)] detector noise, respectively. To make the statistical error comparable, the time of averaging for the solid circles is chosen to be four times longer than the open circles. Both the solid circles and the open circles are in good agreement with the prediction of Eq. (\ref{eq:sol}), shown as the solid line. Provided that the resonator and the detection circuit remain in the linear regime, the predicted values are independent of the presence of thermal and detector noise because both are phase random relative to the periodic driving. Figure \ref{fig:4}(b) shows a similar plot for the ratio $|\langle u^3\rangle /\langle u\rangle ^3|$.  Our measurements support the notion that the deviations of the higher moment ratios $|\langle u^n\rangle /\langle u\rangle ^n|$ from one remain a robust indicator of the presence of frequency noise, even when the thermal noise and/or detector noise is strong.

For all the data presented in this paper, the measurement time is chosen to be much longer than the inverse switching rate, so that the time-averaged values of $u$ and the higher moments are associated with many frequency jumps to reduce the measurement uncertainty. In this limit, the predicted line shapes of the spectra of $|\langle u^2\rangle /\langle u\rangle ^2|$ and $|\langle u^3\rangle /\langle u\rangle ^3|$ are independent of the measurement time. In experiments where thermal noise or detector noise is present, the measurement time may need to be further increased for accurate measurement of these ratios.

\section{COMPARISON OF THE EFFECTS OF TELEGRAPH NOISE AND GAUSSIAN NOISE}

The analysis of the spectra of the higher moments of the complex amplitude is not only limited to telegraph frequency noise. In Figs.$\:$\ref{fig:4}(a) and \ref{fig:4}(b), the dashed line shows the predicted ratios of $|\langle u^2\rangle /\langle u\rangle ^2|$ and $|\langle u^3\rangle /\langle u\rangle ^3|$ when the frequency noise $\xi(t)$ follows a white Gaussian distribution, instead of jumping randomly between two levels \cite{Maizelis2011}. Here, the intensity $D$ of the Gaussian frequency noise is chosen to be $\Delta/2\sqrt{2}$, where $\langle\xi(t)\xi(t')\rangle = 2 D \delta(t-t')$, so that the telegraph noise and Gaussian noise have the same variance. In both Figs.$\:$\ref{fig:4}(a) and \ref{fig:4}(b), the ratios for Gaussian noise deviate from one, but in a manner different from telegraph noise. They remain smaller than one over the entire range of $\delta\omega$, attaining a minimum at $\delta\omega = 0$. In contrast, for the measurements with telegraph noise [Figs.$\:$\ref{fig:2}(b), \ref{fig:2}(d), \ref{fig:4}(a) and \ref{fig:4}(b)], there exists a certain range of $\delta\omega$ where the ratios exceed $1$. It is possible that in addition to determining the existence of frequency noise, the analysis of the higher moments of $u$ might prove useful in the study of noise characteristics. Based on the current study, we cannot conclude if the aforementioned differences in the spectra are specific to the ranges of parameters in our experiment or universal features of telegraph and Gaussian noise. This is an important topic that warrants further theoretical analysis.

\section{CONCLUSIONS}

We have demonstrated that the presence of RTFN can be identified by the moments of the complex amplitude $u$ of an underdamped driven micromechanical oscillator. When either the interval between the two frequencies is large or the switching rate is small, the time-averaged vibration amplitude spectrum exhibits two peaks. They merge with an increasing rate of frequency switching and the spectrum displays an analog of motional narrowing. We also show that the moments of the complex amplitude depend strongly on the frequency noise characteristics. Such analyses of the higher moments of $u$ are particularly useful in revealing the existence of frequency noise when thermal or detector noise is strong. Moreover, this method is directly applicable to other linear resonators with high quality factors, such as Josephson junctions \cite{Grabovskij2012,LiJian2013} and optical cavities.

\section{ACKNOWLEDGMENTS}

We thank M. I. Dykman for useful discussions. This work was supported by a grant from the Research Grant Council of the Hong Kong Special Administrative Region, China (Project No. 600312) and by the NSF through Grant No. CMMI-0856374. F. S. and J. Z. contributed equally to this work.

\begin{appendix}
\section*{APPENDIX: INDEPENDENCE OF THE HIGHER MOMENTS OF COMPLEX VIBRATIONAL AMPLITUDE ON THERMAL NOISE AND DETECTOR NOISE}

Thermal motion of the resonator and/or detector noise in the measurement circuitry often prevents one 
from identifying the existence of frequency noise using conventional methods such as measurement of the ringdown 
of oscillations. In the main text, we demonstrated experimentally that by measuring the ratios of the higher moments
$|\langle u^2\rangle /\langle u\rangle ^2|$ or $|\langle u^3\rangle /\langle u\rangle ^3|$, it is possible
to detect the existence of frequency noise even when the
detector noise is strong, provided that sufficient averaging
is performed. Here, we provide an explanation to why such
an analysis is immune to thermal and/or detector noise.

In the experimental measurement of $u$, the recorded signal 
can be written as:
\begin{equation}
u = (X+\Delta X) + i(Y + \Delta Y).
\end{equation}
$\Delta X$ and $\Delta Y$ originates from thermal 
or detector noise. They have zero mean and are uncorrelated 
with each other and also with $X$ and $Y$ if the resonator and
detection circuit both remain linear. Provided that sufficient
averaging is done, $|\langle u\rangle^n|$ equals
$|\langle X + i Y\rangle^n|$. The numerator of the
ratio considered in the main text, $|\langle u^n\rangle|$, contains cross terms between
$\Delta X$, $\Delta Y$, $X$, and $Y$. For example, consider the case of $n$ = 2 
measured in our experiment:
\begin{align}
&\langle [(X+\Delta X) + i (Y + \Delta Y)]^2\rangle = \langle X^2\rangle - \langle Y^2\rangle 
\nonumber\\
+\,&2(\langle X\rangle \langle \Delta X\rangle-\langle Y\rangle \langle \Delta Y\rangle) +(\langle (\Delta X)^2\rangle - \langle (\Delta Y)^2\rangle)  \nonumber\\
+ \,&i(\langle X\rangle \langle Y\rangle + \langle \Delta X\rangle \langle Y\rangle	+ \langle \Delta Y\rangle \langle X\rangle + \langle \Delta X\rangle \langle \Delta Y\rangle)
\end{align}
All the terms with $\langle \Delta X \rangle$ or $\langle \Delta Y \rangle$ average to zero. 
Because $\Delta X$ and $\Delta Y$ due to thermal or detector noise follow the same distribution,
$\langle (\Delta X)^2\rangle = \langle (\Delta Y)^2\rangle$. Therefore
$|\langle u^2\rangle|$ is identical to the case when thermal or detector
noise is absent. In general, phase noise is incorporated into the 
higher moments while additive noise such as thermal or detector noise 
is not. As a result, with sufficient averaging, the higher moments can 
be used to reveal the existence of frequency noise.
\end{appendix}


\bibliographystyle{apsrev4-1}
%


\end{document}